\documentclass[12pt,preprint]{aastex}
\usepackage{color,soul}
\shorttitle{Spin Alignments in Spiral Pairs}
\shortauthors{Koo \& Lee}
\begin{document}
\title{Detection of the Intrinsic Spin Alignments in Isolated Spiral Pairs}
\author{Hanwool Koo and Jounghun Lee}
\affil{Astronomy program, Department of Physics and Astronomy,
Seoul National University, Seoul  08826, Republic of Korea \\
\email{khw@astro.snu.ac.kr, jounghun@astro.snu.ac.kr}}
\begin{abstract}
An observational evidence for the intrinsic galaxy alignments in isolated spiral pairs is presented. From the catalog of the 
galaxy groups identified by Tempel et al. in the flux limited galaxy sample of the Sloan Digital Sky Survey Data Release 10, 
we select those groups consisting only of two spiral galaxies as isolated spiral pairs and investigate if and how strongly the 
spin axes of their two spiral members are aligned with each other.  We detect a clear signal of intrinsic spin alignment 
in the isolated spiral pairs, which leads to the rejection of the null hypothesis at the $99.9999\%$ confidence level via the 
Rayleigh test. It is also found that those isolated pairs comprising two early-type spiral galaxies exhibit the strongest signal of 
intrinsic spin alignment and that the strength of the alignment signal depends on the angular separation distance as well as 
on the luminosity ratio of the member galaxies. 
Using the dark matter halos consisting of only two subhalos resolved in the EAGLE hydrodynamic simulations, we repeat the 
same analysis but fail to find any alignment tendency between the spin angular momentum vectors of the stellar components 
of the subhalos, which is in tension with the observational result.  A couple of possible sources of this apparent inconsistency 
between the observational and the numerical results are discussed.
\end{abstract}
\section{Introduction}\label{sec:intro}

In the conventional cosmology which aims to track down the expansion and growth history of the whole universe \citep{linder05}, 
it is generally believed that the deeper we look into the past of the universe, the more cosmological information we would 
obtain about the universe.  Quite true as this belief may be, the conventional cosmology is inevitably hindered by unknown 
(but likely existent) systematics involved with the observations of distant objects.  For instance,  it has been 
recently warned that the luminosity-distance relation of the distant type Ia supernovae (SNIa) which is regarded as the 
first direct evidence for the cosmic acceleration \citep{SN98,SN99} may be substantially contaminated by the 
systematics involved in the conversion of the brightness of SNIa into its intrinsic luminosity \citep{nielsen-etal16}. 

For the past two decades,  we have witnessed the emergence of a "near-field cosmology", which utilizes the physical 
properties of the local objects as a test statistics \citep{BP06}.  Although the evolution history of the universe cannot be 
traced by observing the local objects, it was realized that the initial conditions of the universe may be imprinted 
as fossil records in the local objects and thus can be inferred through careful analyses of the physical properties of 
the nearby objects \citep{BP06}.  The obvious merit of using the near-field diagnostics is nothing but relatively low 
uncertainties in the observations of the local objects, which will allow us to uncover the imprinted (and likely vagrant) fossil 
records left from the early universe. Furthermore,  the near-field statistics will help us undertake a delicately difficult mission of 
discriminating viable alternative cosmologies from the standard $\Lambda$CDM model where the cosmological constant 
$\Lambda$ fuels the cosmic acceleration and the most dominant matter content is cold dark matter (CDM). 
 
A variety of near-field diagnostics has so far been developed, which includes the environmental link of the dynamic-to-lensing mass 
ratio of galaxy clusters, turn-around radii of the galaxy groups, emptiness of cosmic voids, dynamics of the Local Group, correlations 
of the galaxy peculiar velocities, luminosity-to-mass ratio of isolated dwarf galaxies and so forth 
\citep[][and references therein]{zhao-etal11,PT14,ade-etal17,carlesi-etal17,hut-etal17,pee17}. 
The intrinsic spin alignments in the galactic pairs is one of those near-field diagnostics suggested by \citet{lee12} who claimed 
that it would be useful not only to test the standard $\Lambda$CDM cosmology but also to distinguish between different coupled 
dark energy (cDE) models in which the scalar field dark energy is coupled to dark matter particles 
\citep[for a review][and references therein]{AT10}. 

Analyzing the isolated halos consisting of only two subhalos from the adiabatic hydrodynamic simulations (H-CoDECS) that 
were run for the cDE models as well as for the $\Lambda$CDM cosmology \citep{baldi-etal10,codecs},  \citet{lee12} explored 
whether or not the directions of the three dimensional (3D) spin vectors of two subhalos are aligned with each other and found 
that some cDE models produced a statistically significant signal of spin alignments in the isolated subhalo pairs while 
the $\Lambda$CDM case yielded no signal. 
Noting that the strength of the spin alignment in the isolated subhalo pairs depends on the strength of the DE-DM coupling as 
well as on the shapes of the DE potentials, \citet{lee12} suggested  that the spin alignments in the isolated galactic pairs 
could provide a powerful local diagnostics to test the dark sector coupling. 

As mentioned in  \citet{lee12}, however, in order to verify the practical usefulness of this new local diagnostics, two additional 
tasks must be performed.  The first task is to make a theoretical prediction based on a fully hydrodynamical simulation that is 
capable of resolving luminous galaxies. What \citet{lee12} measured was the alignments of the spin axes not of the luminous galaxies 
but of the DM subhalos resolved in the H-CoDECS which was not capable of simulating the formations of stars and galaxies 
\citep{codecs}. Given that the spin axes of the luminous galaxies are not necessarily aligned with those of their underlying DM particles 
\citep[e.g.,][]{hah-etal10,ten-etal17,ZS17},  the numerical result of \citet{lee12} cannot be directly compared with those from the 
observations. 

The second task is to improve the observational analyses.  Although several works already investigated the spin alignments 
of the isolated galactic pairs \citep[e.g.,][]{GT78,SLW79,hel84,oos93,PC04,cer-etal10,lee12},  
it is still inconclusive whether the intrinsic signals truly exist or not.
The main reason for having failed to confirm or to refute convincingly the existence of the spin alignments in the isolated galactic 
pairs is the poor approximations that the previous works relied on for their analyses. For instance, \citet{cer-etal10} approximated 
the alignment angles between the spin axes of the member galaxies by the differences in their position angles without 
attempting to measure directly the 3D spin vectors.  

Besides, in the previous works, an isolated galaxy pair was selected not as a bound system composed of two galaxies without 
being embedded in a larger structure but just as two galaxies closely located with each other. 
However, the mechanism which generates the alignments between the spin axes of the closely located neighbor 
galaxies is different from what is responsible for the alignments in the isolated galaxy pairs. 
While the former is usually ascribed to the combined effects of the spatial correlations of the 
surrounding tidal field \citep{pee69,dor70,whi84,dub92,BE87,CT96,LP00,LP01,por-etal02} and the filamentary merging along the 
cosmic web \citep{codis-etal12},  
the latter is believed to be generated by nonlinear galaxy-galaxy interaction \citep{GT78,hel84}.
It should be necessary to sort out properly isolated galaxy pairs that forms a bound system, to properly measure the alignment 
angles directly from the 3D spin axes, and then to employ a robust statistical test of the null hypothesis of no alignment. 

Here, we intend to perform the above two tasks. The main contents of the next three Sections 
can be summarized as follows. In Section \ref{sec:ob}, we present an observational evidence for the existence of the intrinsic spin 
alignments in the isolated spiral pairs. In Section \ref{sec:num}, we show that the numerical analysis based on a full hydrodynamical 
simulations fails to detect any signal of intrinsic spin alignment in the isolated pair system. In Section \ref{sec:dis}, we discuss 
possible origins of the inconsistencies between the numerical and the observational results. 

\section{An Observational Signal of Intrinsic Spin Alignments in the Isolated Pairs}\label{sec:ob}

\citet{tempel-etal14} identified the galaxy groups by applying a modified friends-of-friends (FoF) group-finding algorithm 
to the flux-limited sample of the galaxies from the tenth data release of the Sloan Digital Sky \citep[SDSS DR10][]{sdssdr10}.  
According to them, the modification of the FoF group finder was made to accommodate the finger-of-god effect on the radial 
distances to the SDSS DR10 galaxies.  The group catalog of \citet{tempel-etal14} provides various information on the properties 
of the member galaxies belonging to each group such as their galaxy identification number (ID), redshifts, equatorial positions, 
morphologies, absolute magnitudes, and so forth.

The morphological type of each member galaxy was obtained from the Galaxy Zoo Project of \citet{zoo} and from the Bayesian 
analysis of \citet{hue-etal11}. The former classified the morphological types ($T_{z}$) of the galaxies as unclear ($T_{z}=0$), 
spiral ($T_{z}=1$) and elliptical ($T_{z}=2$). Meanwhile, the latter estimated the four  Bayesian probabilities, 
$P(E),\ P(S0),\ P(Sab),\ P(Scd)$, of each Sloan galaxy being classified as an elliptical (E), lenticular (S0), early-type spiral 
(Sab), and late-type spiral (Scd), respectively.  

From the FoF group catalog of \citet{tempel-etal14}, we search for the isolated spiral pairs which satisfy the following three 
conditions:  (1) $n_{g}=2$ where $n_{g}$ is the number of the member galaxies constituting a FoF group; (2)  
$P_{\rm x} = P(Scd)$ or $P_{\rm x}=P(Sab)$ where $P_{\rm x}\equiv {\rm max}\{P(E),\ P(S0),\ P(Sab),\ P(Scd)\}$; 
(3) $T_{z} = 1$. An {\it isolated pair} means a FoF group composed of only two members without being embedded in any 
other larger FoF groups.  
From here on, the Sab galaxies refer to those spiral galaxies satisfying the conditions of $T_{z}=1$ and $P_{\rm x}=P(Sab)$, 
while the Scd galaxies to those satisfying $T_{z}=1$ and $P_{\rm x}=P(Sab)$.

\begin{figure}
\begin{center}
\includegraphics[scale=0.8]{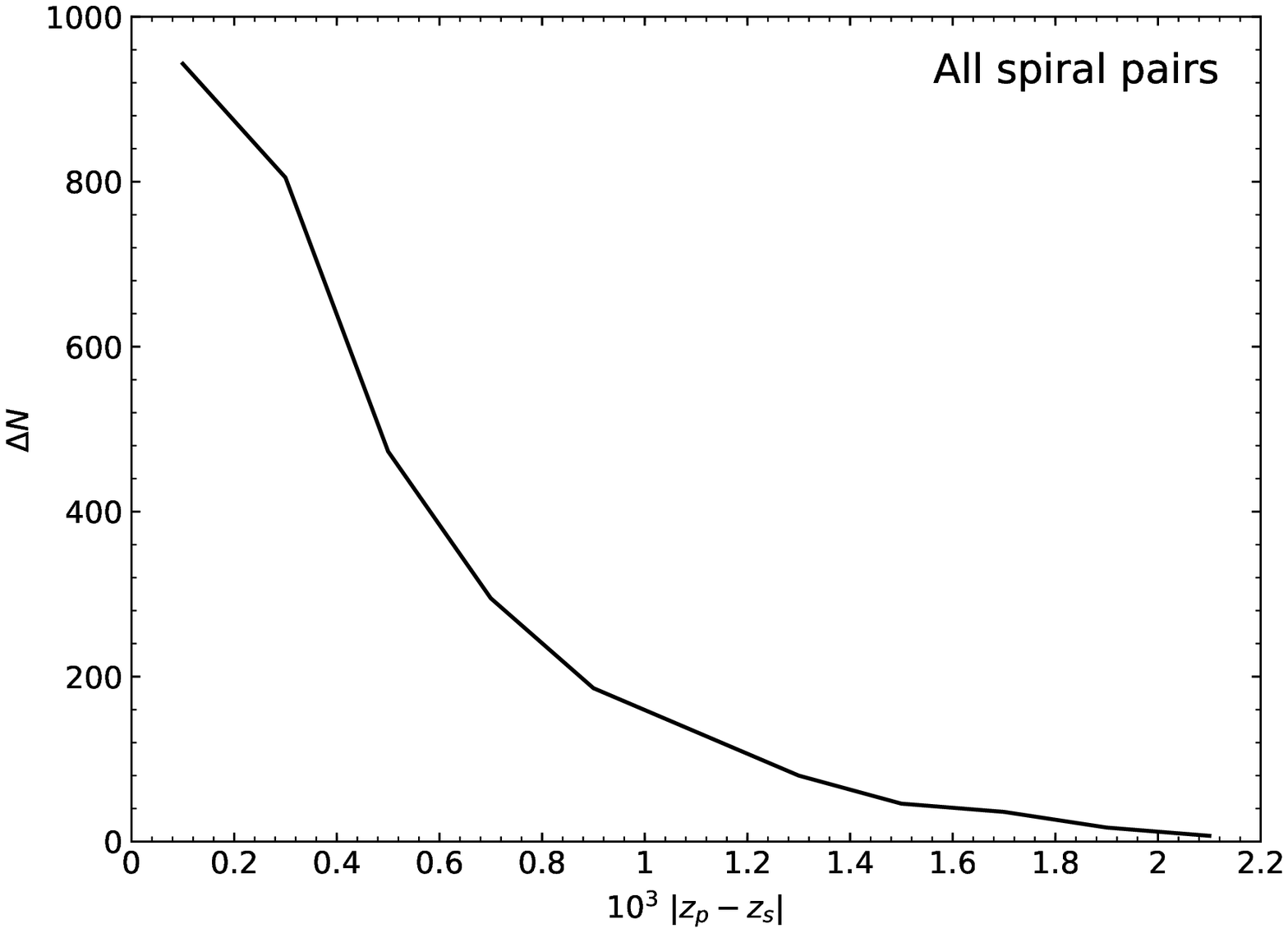}
\caption{Distribution of the numbers of the isolated spiral pairs versus the differences between the redshifts 
of the primary and the secondary member galaxies.}
\label{fig:delz}
\end{center}
\end{figure}

We end up having a total of $3021$ isolated spiral pairs from the FoF group catalog of \citet{tempel-etal14}.  For each isolated 
spiral pair, we compare the absolute $r$-band magnitudes of its two members with each other and define a primary (secondary) 
member as the one with larger (smaller) values of $\vert M_{r}\vert$.  The first two columns of Table \ref{tab:ob} list the range of the 
redshifts and the absolute $r$-band magnitudes of the member galaxies in the isolated spiral pairs. 
Figure \ref{fig:delz} shows the number distribution of the redshift difference, $\vert z_{p}-z_{s}\vert$ between the primary and the 
secondary members of the selected isolated spiral pairs. 
As can be seen, the majority of the pairs has $\vert z_{p}-z_{s}\vert \le 0.001$. 

As in \citet{lee11},  we adopt the {\it thin circular disk approximation} to estimate the directions of the spin angular momentum 
vectors of two members of each isolated spiral pair, which requires such information as the equatorial positions (RA and Dec), 
position angles ($\psi_{p}$), axial ratios ($R_{a}$) and morphological types of the member galaxies.   
Although the FoF group catalog of \citet{tempel-etal14} does not contain information on $\psi_{p}$ and $R_{a}$, we extract the 
values of $\psi_{p}$ and $R_{a}$ of each member galaxy 
from the web-site of the SDSS DR10\footnote{http://www.sdss3.org/dr10/} by matching the galaxy ID's. 

\begin{figure}
\begin{center}
\includegraphics[scale=0.3]{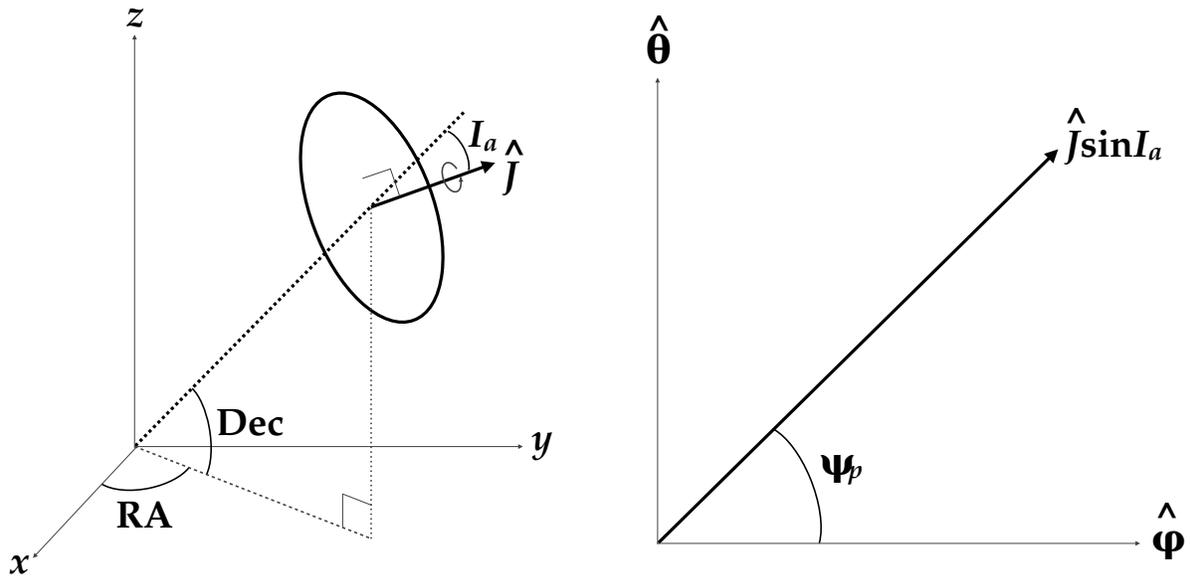}
\caption{Configuration of the spin angular momentum vector of a thin circular disk-like spiral galaxy in the 
equatorial coordinate system.}
\label{fig:spin_con}
\end{center}
\end{figure}

For the convenience of the readers, we briefly review the thin circular disk approximation: 
Let $\hat{\bf J}$ denote a unit vector in the direction of the spin axis of a member galaxy in a given isolated spiral pair. 
Expressing $\hat{\bf J}$ in the  spherical polar coordinates such as $\hat{\bf J}=(\hat{J}_{r}, \hat{J}_{\theta}, \hat{J}_{\phi})$, 
one can say that its radial component, $\hat{J}_{r}$, is the projection of $\hat{\bf J}$ onto the direction of the line-of-sight toward 
the spiral galaxy while its polar and azimuthal coordinates, $\hat{J}_{\theta}$ and $\hat{J}_{\phi}$, are the projection onto the plane 
orthogonal to the line of sight direction.  
Under the assumption that a spiral galaxy is a thin circular disk and its spin axis is perpendicular to the disk plane, the observed 
axial ratio, $R_{a}$, of a spiral galaxy should be the same as the cosine of the inclination angle, which is in turn equal to 
the {\it magnitude} of the radial coordinate of the spin vector: 
$\vert\hat{ J}_{r}\vert\approx \cos I_{a} = R_{a}$ \citep[see][and references therein]{pen-etal00}. 
Additional information on the position angle, $\psi_{p}$, of a spiral galaxy determines the other two coordinates: 
$\hat{J}_{\theta} = \sqrt{1 - \hat{J}^{2}_{r}}\sin\psi_{p}$ and 
$\hat{J}_{\phi} = \sqrt{1 - \hat{J}^{2}_{r}}\cos\psi_{p}$ (see Figure \ref{fig:spin_con}).

This simple routine based on the thin circular disk approximation for the determination of $\hat{\bf J}$ has two downsides. 
First, it is blind to the sign of the radial component $\hat{J}_{r}$ \citep{pen-etal00}.  
Second, it's validity depends sensitively on the thinness of a disk. In other words, for a real spiral galaxy which usually have a  
central bulge, this routine is likely to fail.  While the first weak point is a rather generic one that cannot be avoided, the second 
one may be overcome by making the following modification in the relation between $I_{a}$ and $R_{a}$, as \citet{HG84} 
suggested: 
\begin{equation}
\label{eqn:I_f}
\hat{J}_{r}\cos^{2}I_{a}\approx (R^{2}_{a} - I^{2}_{f})/(1 - I^{2}_{f})\, ,
\end{equation}
where $I_{f}$ is a {\it flatness parameter} introduced by \citet{HG84} to account for the deviation of a spiral galaxy with a thick bulge 
from a thin disk.  Following the suggestion of \citet{HG84}, we set the values of the flatness parameter at $I_{f}=0.1$ and $I_{f}=0.2$ 
for the Sab and Scd galaxies, respectively.  

\begin{figure}
\begin{center}
\includegraphics[scale=0.4]{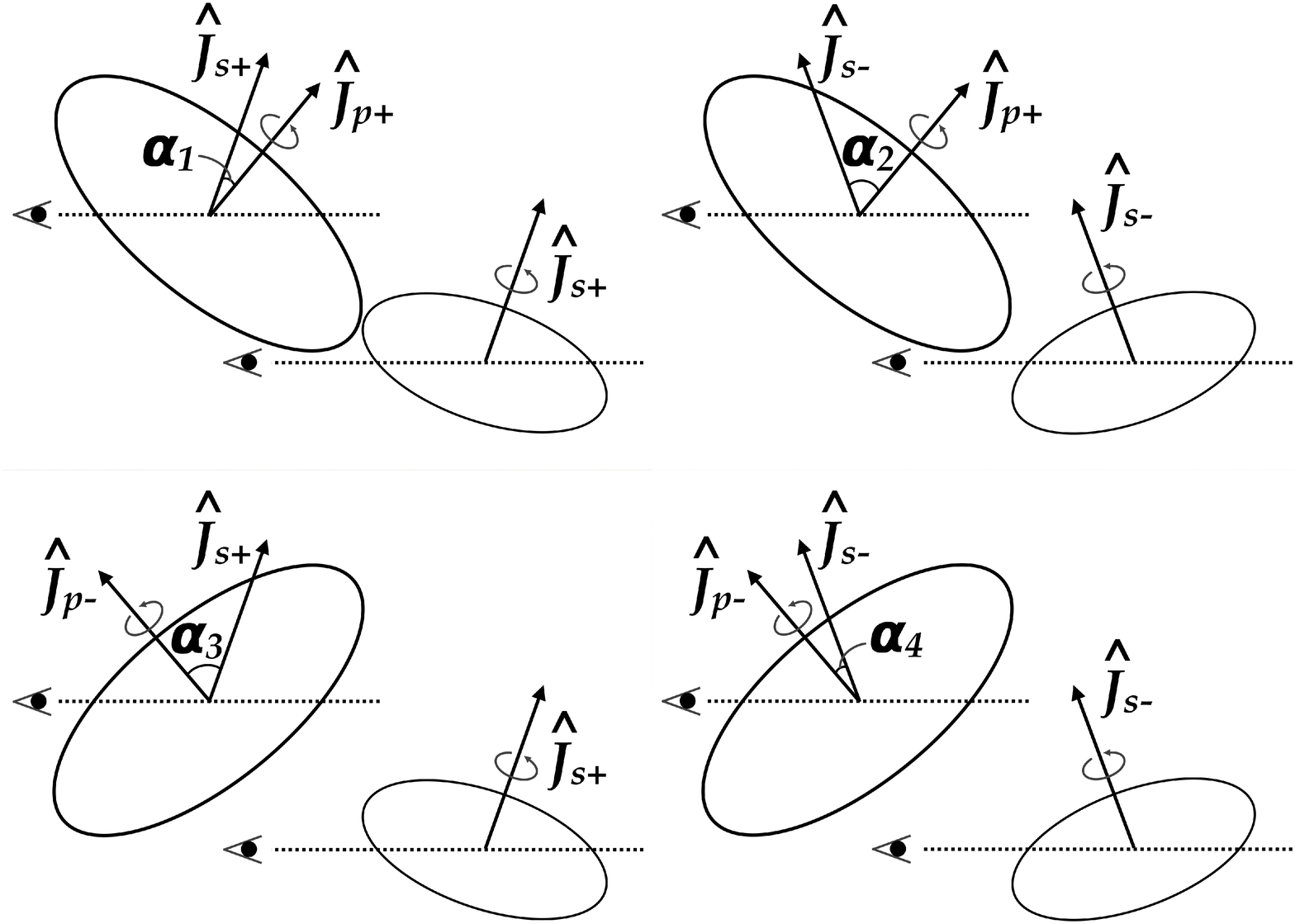}
\caption{Illustration of the four-fold degeneracy involved with the measurement of the cosine of the 
alignment angles between the spin axes of the member galaxies in a spiral pair.}
\label{fig:spin_deg}
\end{center}
\end{figure}

If the radial component of $\hat{\bf J}$ is positive, then the Cartesian coordinates, 
$\hat{\bf J} = (\hat{J}_{x}, \hat{J}_{y}, \hat{J}_{z})$, of the unit spin vector of a spiral galaxy can now be written as 
\citep{pen-etal00}: 
\begin{eqnarray}
\hat{J}_{x+}&=&+\hat{J}_{r}\sin\theta\cos\phi + \hat{J}_{\theta}\cos\theta\cos\phi  - \hat{J}_{\phi}\sin\phi\, ,\\
\hat{J}_{y+} &=&+\hat{J}_{r}\sin\theta\sin\phi  + \hat{J}_{\theta}\cos\theta\sin\phi + \hat{J}_{\phi}\cos\phi\, , \\ 
\hat{J}_{z+} &=&+\hat{J}_{r}\cos\theta - \hat{J}_{\theta}\sin\theta\, ,
\end{eqnarray}
while, for the case of the negative radial component, the sign in front of $\hat{J}_{r}$ should be reversed as
\begin{eqnarray}
\hat{J}_{x-} &=&-\hat{J}_{r}\sin\theta\cos\phi + \hat{J}_{\theta}\cos\theta\cos\phi  - \hat{J}_{\phi}\sin\phi\, ,\\
\hat{J}_{y-} &=&-\hat{J}_{r}\sin\theta\sin\phi  + \hat{J}_{\theta}\cos\theta\sin\phi + \hat{J}_{\phi}\cos\phi\, ,\\
\hat{J}_{z-} &=&-\hat{J}_{r}\cos\theta - \hat{J}_{\theta}\sin\theta\, . 
\end{eqnarray}
The sign ambiguity in the determination of $\hat{\bf J}$ leads us to have four different realizations 
(say, $\alpha_{1},\ \alpha_{2},\alpha_{3},\ \alpha_{4})$ of the alignment angles 
between the spin axes of the primary and the secondary member galaxies in each isolated spiral pair, 
as shown in Figure \ref{fig:spin_deg} \citep[see discussions in][]{pen-etal00,lee11}: 
\begin{eqnarray}
\label{eqn:cost1}
\cos\alpha_{1} &=& \vert\hat{\bf J}_{p+}\cdot\hat{\bf J}_{s+}\vert\, , \\
\label{eqn:cost2}
\cos\alpha_{2} &=& \vert\hat{\bf J}_{p+}\cdot\hat{\bf J}_{s-}\vert\, , \\ 
\label{eqn:cost3}
\cos\alpha_{3} &=& \vert\hat{\bf J}_{p-}\cdot\hat{\bf J}_{s+}\vert\, , \\ 
\label{eqn:cost4}
\cos\alpha_{4} &=& \vert\hat{\bf J}_{p-}\cdot\hat{\bf J}_{s-}\vert\, .
\end{eqnarray}
From a total of $N_{sp}$ isolated spiral pairs, we have $4N_{sp}$ realizations of $\cos\alpha$.

\begin{figure}
\begin{center}
\includegraphics[scale=0.8]{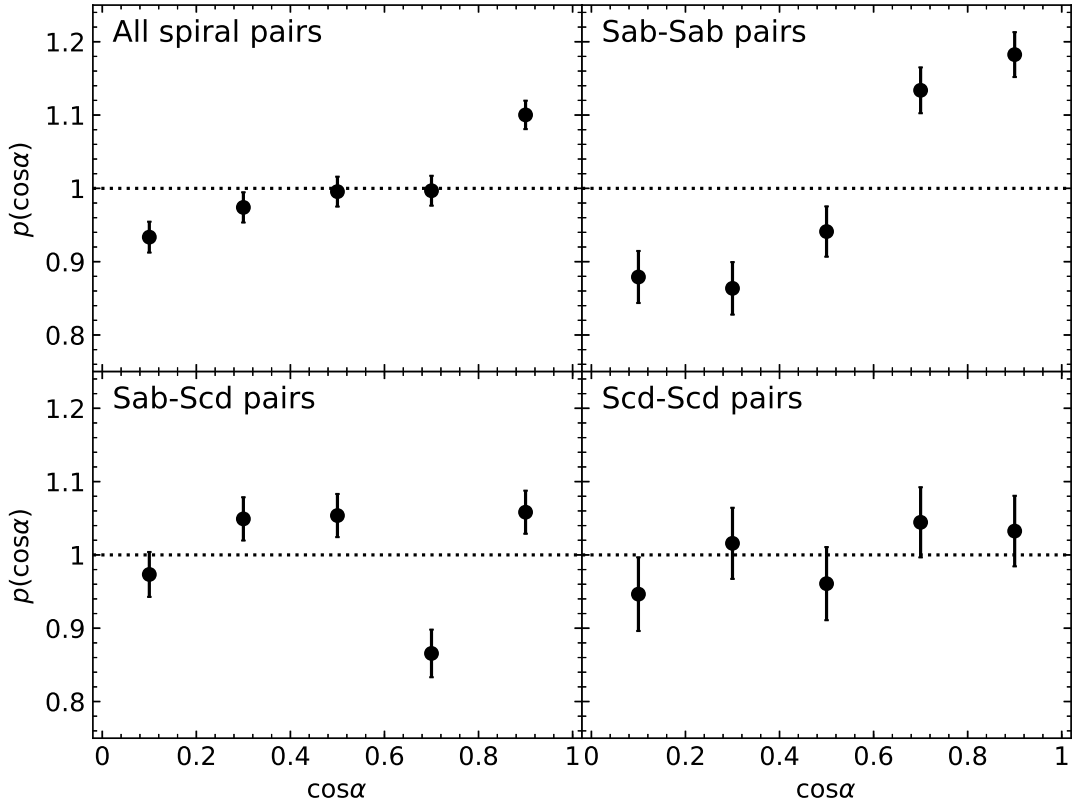}
\caption{Probability density functions of the cosines of the angles between the spin axes of the members 
in the isolated spiral pairs from the SDSS DR10: all types, early, mixed, late-type spiral pairs in the top-left, top-right, 
bottom-left and bottom-right panels, respectively.  In each panel, the dotted line displays a uniform distribution.}
\label{fig:pcosa}
\end{center}
\end{figure}

Partitioning the range of $[0,\ 1]$ into five small intervals each of which has an equal length of $\Delta\cos\alpha = 0.2$, 
we calculate $p(\cos\alpha_{i})$ at each interval of $[\cos\alpha_{i}, \cos\alpha_{i} + \Delta\cos\alpha_{i}]$ as 
$p(\cos\alpha_{i}) = n_{i, sp}/(\Delta\cos\alpha\cdot 4N_{sp})$, where $n_{i, sp}$ 
denotes the number of the realizations of $\cos\alpha$ belonging to the $i$th interval of $\cos\alpha$. 
The top-left panel of Figure \ref{fig:pcosa} displays as filled circles the probability density function of the cosines of 
the angles of the isolated spiral pairs, $p(\cos\alpha)$, from the flux-limited samples of the SDSS DR10. The errors are 
Poissonian calculated as $1/\sqrt{(n_{i,sp}-1)}$. 
If there were no alignment between the spin vectors of the two members in the isolated spiral pairs, then the expectation of the 
probability density function would be constant as $p(\cos\alpha)=1$ (dotted line).  As shown, the probability density function 
exhibits an obvious tendency to increase with the increment of $\cos\alpha$, revealing that the directions of the spin axes of the 
primary and the secondary members in the isolated spiral pairs are not random but preferentially aligned with each other.


To see if the strength of the intrinsic spin alignment depends on the morphological types of the spiral 
galaxies,  we divide the isolated spiral pairs into three different subsamples: Sab-Sab, Scd-Scd and Sab-Scd pairs. 
The first (second) subsample consists only of two early-type (two late-type) galaxies, while the third sample contains those 
pairs consisting of one Sab and one Scd galaxies. For each of the three subsamples, we redo all of the calculations described 
in the above. The numbers of the Sab-Sab, Sab-Scd and Scd-Scd pairs are listed in the last three columns of Table \ref{tab:ob}, 
respectively.
The top-right, bottom-left, and bottom-right panels of Figure \ref{fig:pcosa} show $p(\cos\alpha)$ versus $\cos\alpha$ 
as filled circles for the three cases of the Sab-Sab, Sab-Scd and Scd-Scd pairs, respectively. 
As shown, the Sab-Sab pairs exhibit a clear signal of intrinsic spin alignment, while no statistically significant 
signals are found from the Sab-Scd nor from the Scd-Scd pairs. For the case of the Sab-Scd pairs, the 
probability density function $p(\cos\alpha)$ shows an abrupt drop at the fourth bin from the left.  

Any systematics in our measurements of the alignment angles could cause spurious correlations between the values of 
$p(\cos\alpha)$ at two adjacent $\cos\alpha$-intervals. Concerning about the existence of this possible correlations, we 
calculate the generalized chi-squared as 
\begin{equation}
\label{eqn:chi2}
\chi^{2} = \sum_{i=1}^{5} \left[p(\cos\alpha_{i})-1\right]C_{ij}\left[p(\cos\alpha_{j})-1\right]\, ,
\end{equation}
where the covariance matrix ${\bf C}=(C_{ij})$ is calculated from the $10000$ Bootstrap resamples as 
\begin{equation}
\label{eqn:cov}
C_{ij}= \frac{1}{N_{boot}}\sum_{k=1}^{N_{boot}}
\left[p^{k}(\cos\alpha_{i})-p(\cos\alpha_{i})\right]\left[p^{k}(\cos\alpha_{j})-p(\cos\alpha_{j})\right]\, .
\end{equation}
where $p^{k}(\cos\alpha_{i})$ is the probability density function from the $k$th Bootstrap resample and $N_{boot}=10000$ is the 
total number of the Bootstrap resamples. 
The subtraction of unity from $p^{k}(\cos\alpha_{i})$ in Equation (\ref{eqn:cov}) is in accordance with the fact that the null hypothesis 
predicts $\langle p(\cos\alpha_{i})\rangle=1$ at each $\cos\alpha$-interval. This generalized chi-squared statistics reject the null 
hypothesis at the confidence levels of $98.37\%$ and $99.99\%$ for the cases of all spiral and Sab-Sab pairs, respectively.

\begin{figure}
\begin{center}
\includegraphics[scale=0.8]{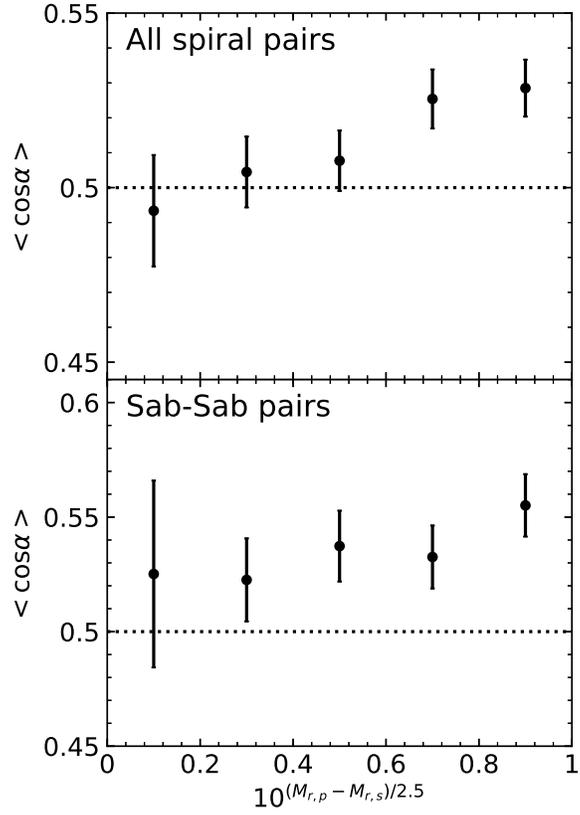}
\caption{Mean values of the cosines of the alignments angles versus the luminosity ratios of the member 
galaxies in the isolated spiral pairs.}
\label{fig:mean_md}
\end{center}
\end{figure}

We investigate the dependence of the strength of intrinsic spin alignments on the luminosity ratios by calculating 
$\langle\cos\alpha\rangle$ as a function of $10^{(M_{r,p}-M_{r,s})/2.5}$, 
where $M_{r,p}$ and $M_{r,s}$ denote the $r$-band absolute magnitudes of the primary and secondary galaxies in each 
isolated spiral pairs, respectively. Dividing the whole range $10^{(M_{r,p}-M_{r,s})/2.5}$ into several short intervals of 
equal length, we evaluate the mean value of $\cos\alpha$ averaged over those pairs whose luminosity ratios lie in 
the range of a given interval. The results are shown in the top and bottom panels of Figure \ref{fig:mean_md} for the cases of 
all spiral and Sab-Sab pairs, respectively.  As can be seen, for both of the cases, the alignment signals tend to be stronger as 
the luminosity ratios are closer to unity (i.e., $10^{(M_{r,p}-M_{r,s})/2.5}\approx 1$). This result implies that those isolated spiral 
pairs with members having more comparable masses yield stronger signals of spin alignments. 

\begin{figure}
\begin{center}
\includegraphics[scale=0.8]{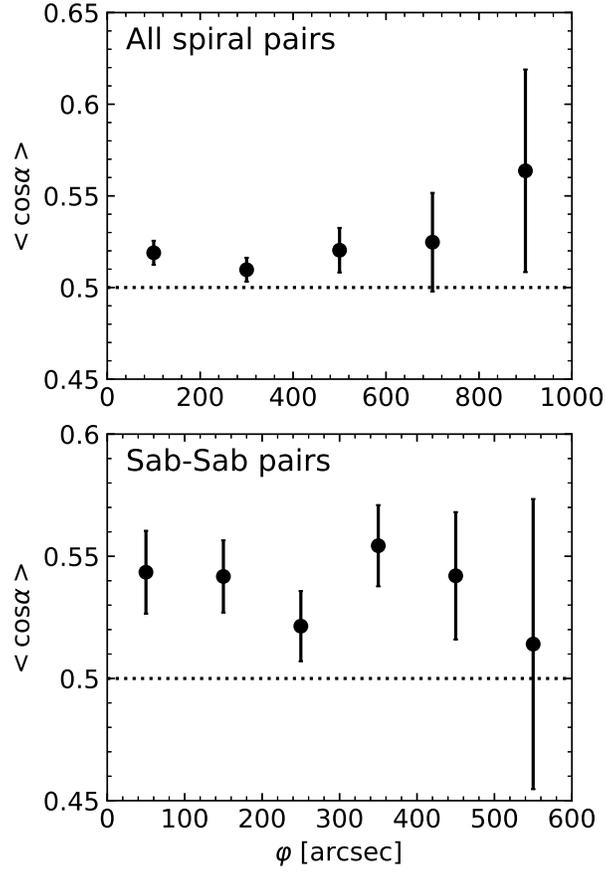}
\caption{Mean values of the cosines of the alignments angles versus the angular separation distance 
between the member galaxies in the isolated spiral pairs.}
\label{fig:mean_sep}
\end{center}
\end{figure}

The dependence of the strength of intrinsic alignment on the angular separation distances between the primary and the 
secondary galaxies in the isolated spiral pairs is also investigated in a similar manner by measuring $\langle\cos\alpha\rangle$ 
versus the angular separation distance, $\varphi$, which can be readily calculated from the equatorial coordinates of the 
member galaxies:
\begin{equation}
\cos\varphi = \cos({\rm Dec}_{p})\cos({\rm Dec}_{s}) - \sin({\rm Dec}_{p})\sin({\rm Dec}_{s})\cos({\rm RA}_{p}-{\rm RA}_{s})\, ,
\end{equation}
where $\{{\rm Dec}_{p},\ {\rm RA}_{p}\}$ ($\{{\rm Dec}_{s},\ {\rm RA}_{s}\}$) is the equatorial coordinates of the primary 
(secondary) members in each isolated spiral pair. Note that this 2D projected distance between the two members in each 
isolated spiral pair can be obtained without specifying the background cosmology. 
The top and bottom panels of Figure \ref{fig:mean_sep} show $\langle\cos\alpha\rangle$ versus $\varphi$ for the cases of 
all spiral and Sab-Sab pairs, respectively. 
As can be seen, the intrinsic spin alignments in the isolated spiral pairs show a mild tendency to increase with the decrement of the 
angular separation distances: The more closely located the member galaxies are, the more strongly their spin axes are aligned 
with each other.

So far, we have relied on the {\it parametric} chi-squared statistics to find the non-uniformity in the distributions of the alignment angles 
between the 3D spin axes of the spiral galaxies in the isolated pairs.  The confidence level, $98.37\%$, at which the chi-squared test 
rejects the null hypothesis for the case of all spiral pairs, however, is not high enough to confirm the detection of the non-uniformity.  
To complement the chi-squared statistics, we perform a {\it non-parametric} Rayleigh test operating directly on the alignment angles 
\citep{FB12}.  For each isolated pair, we project the 3D unit spin vectors of the primary and the secondary spiral galaxies 
($\hat{\bf J}_{p}$ and $\hat{\bf J}_{s}$, respectively) onto the plane of the sky perpendicular to the line-of-sight direction, $\hat{\bf r}_{p}$ 
to obtain the corresponding 2D spin vectors ($\hat{\bf j}_{p}$ and $\hat{\bf j}_{s}$, respectively)
\begin{equation}
\label{eqn:hatj}
\hat{\bf j}_{\gamma}=\frac{\hat{\bf J}_{\gamma} - (\hat{\bf J}_{\gamma}\cdot\hat{\bf r}_{p})\hat{\bf r}_{p}}{\vert\hat{\bf J}_{\gamma} 
- (\hat{\bf J}_{\gamma}\cdot\hat{\bf r}_{p})\hat{\bf r}_{p}\vert},\ \qquad {\rm for}\quad \gamma = {\rm p},\ {\rm s}\, , 
\end{equation}
and then calculate the alignment angles, $\beta$, as
\begin{eqnarray}
\label{eqn:2dangle}
\beta=\left\{
        \begin{array}{ll}
            +\cos^{-1}{(\hat{\bf j}_{p}\cdot\hat{\bf j}_{s})}\, & {\rm if}\quad (\hat{\bf j}_{p}\times\hat{\bf j}_{s})\cdot\hat{\bf r}_{p} < 0\, , \\
            -\cos^{-1}{(\hat{\bf j}_{p}\cdot\hat{\bf j}_{s})}\, & {\rm if}\quad (\hat{\bf j}_{p}\times\hat{\bf j}_{s})\cdot\hat{\bf r}_{p} \geq 0\, .
        \end{array}
    \right.
\end{eqnarray}

\begin{figure}
\begin{center}
\includegraphics[scale=0.8]{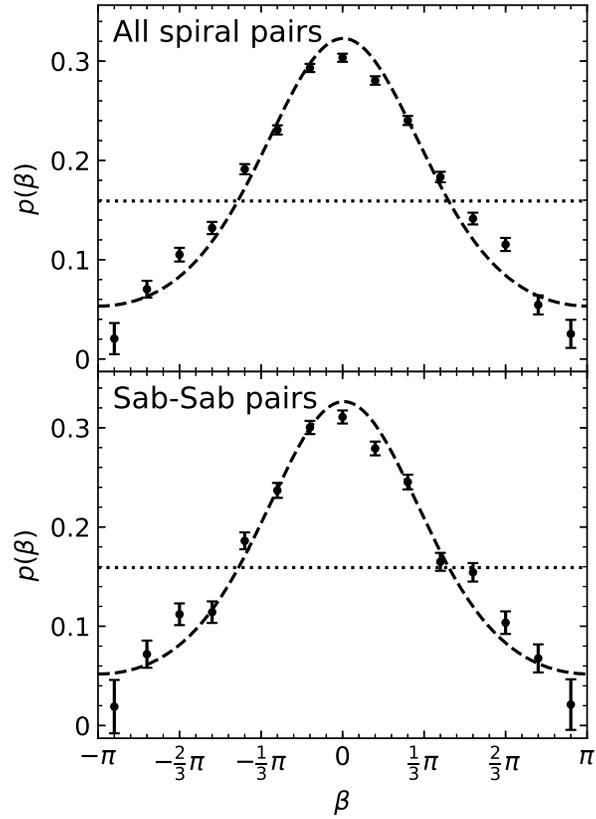}
\caption{Probability density functions of the alignment angles defined in the plane of sky between the 2D projected spin vectors of 
theprimary and the secondary galaxies for the case of all spiral (top panel) and the Sab-Sab pairs (bottom panel). 
The best-fit von Mises (dashed lines) and the uniform distribution (dotted line) are also shown for comparison 
in each panel.}
\label{fig:pbeta}
\end{center}
\end{figure}

Dividing the full range, $[-\pi,\ \pi]$, of $\beta$,  into a dozen of short intervals each of which has an equal length of 
$\Delta\beta = 2\pi/15$, we determine $p(\beta_{i})$ at each interval of 
$[\beta_{i}, \beta_{i} + \Delta\beta_{i}]$ as $p(\beta_{i}) = n_{i, sp}/(\Delta\beta\cdot 4N_{sp})$, 
where $n_{i, sp}$ denotes the number of the realizations of $\beta$ belonging to the $i$th interval of $\beta$. 
The top and bottom panels of Figure \ref{fig:pbeta} displays the probability density functions of the alignment angles, $p(\beta)$, 
(solid circular dots) along with the Poisson errors, $\Delta p(\beta_{i}) = 1/(2\pi\sqrt{n_{i,sp}-1})$ for the cases of all spiral pairs and of 
the Sab-Sab pairs, respectively.  As shown, the observational results significantly deviate from the circular uniform distributions 
$p(\beta)=1/2\pi$ (dotted lines).  It is found that the Rayleigh test rejects the null hypothesis of no alignment at the 
confidence levels as high as $99.9999\%$ for both of the cases. 

Deviation of $p(\beta)$ from the uniform distribution and its Gaussian-like shape spurs us to fit it to the following von Mises 
distribution characterized by two characteristic parameters, $\mu$ an $\kappa(>0)$ \citep{briggs93,fisher93,pew-etal13}: 
\begin{equation}
\label{eqn:von}
p(\beta\ \vert\ \mu,\kappa)=\frac{e^{\kappa\cos(\beta-\mu)}}{2\pi I_{0}(\kappa)}\, ,
\end{equation}
where $I_{0}(\kappa)$ is the modified Bessel function of the first kind of order $0$. The two parameters, $\mu$ and $1/\kappa$, are to 
the von Mises distributions what the mean and the variance are to the Gaussian distribution. The value of $\mu$ represents 
the location where the probability density function, $p(\beta)$, reaches its maximum, while $1/\kappa$ measures the width of 
the distribution. The von Mises distribution will be reduced to the uniform distribution provided $\kappa=0$.  
The larger the value of $\kappa$ is, the narrower shape the von Mises distribution will have.

The best-fit values of $\mu$ and $\kappa$ for the cases of all spiral and Sab-Sab pairs are found by the maximum likelihood method and 
listed in Table \ref{tab:best_fit}. As can be read, the best-fit values of $\mu$ are effectively zero for both of the cases, which implies that 
the 2D spin vectors of the primary and the secondary galaxies in the isolated pairs are indeed preferentially aligned with each other. 
The significant deviation of the best-fit values of $\kappa$ from zero also confirms our finding of non-uniformity in the probability 
density function of $p(\beta)$. The von Mises distributions with these best-fit parameters are plotted as dashed lines in 
Figure \ref{fig:pbeta}. As can be seen, the overall shapes of the probability density function, $p(\beta)$, indeed, match well the von Mises 
distributions with the best-fit parameters, for both of the cases. 

\section{Numerical Predictions of the $\Lambda$CDM Cosmology}\label{sec:num}

To compare the observational results obtained in Section \ref{sec:ob} with the theoretical prediction of the $\Lambda$CDM 
cosmology, we utilize the data from the EAGLE project of cosmological hydrodynamic simulations that were run on a periodic 
box of $100$ comoving mega parsecs (cMpc) with a total of $2\times 1504^{3}$ dark matter particles 
\citep{eagle_subgrid,eagle_catalog,eagle}.  
Assuming a $\Lambda$CDM universe with initial conditions constrained by the angular power spectrum of the temperature 
fluctuation field of the Cosmic Microwave Background (CMB) from the Planck survey \citep{planck14a}, the EAGLE project 
computed the influences of various baryonic processes as well as gravity to simulate the realistic evolutions of the luminous 
galaxies that reside in the DM halos. \citep{eagle}.  The standard FoF group finder and the SUBFIND algorithms 
\citep{subfind,eagle_catalog} were used by the EAGLE project to find DM halos and to resolve the subhalos within 
the virial radius of each DM halo, respectively. 
From the website of the EAGLE hydrodynamic simulations\footnote{http://icc.dur.ac.uk/Eagle/database.php},  one can access the 
catalogs of the FoF groups and their subhalos at various redshifts from $z=127$ to $z=0$ and extract such information as the spin 
angular momentum vectors, position vectors and virial masses of the subhalos. 

We consider two different redshifts, $z=0$ and $0.1$, to match the redshift ranges of the isolated spiral pairs analyzed in 
Section \ref{sec:ob}.  We use the following criteria to select a halo as an isolated pair system from the DM halo catalog: 
(i) $N_{\rm sub}=2$; (ii) $N_{t}\ge N_{t,c}=1000$; (iii) $N_{s}\ge N_{s,c}=100$, where $N_{\rm sub}$ is the number of the subhalos
contained in a host halo, $N_{t}$ and $N_{s}$ are the numbers of the constituent particles of a host halo and its secondary subhalo, 
respectively,  while $N_{t,c}$ and $N_{s,c}$ represent the thresholds on the values of $N_{t}$ and $N_{s}$, respectively. 
The primary and the secondary subhalos of a host halo with $N_{\rm sub}=2$ are determined by their virial masses. 
The second and third criteria are required to reduce possible systematics caused by the low resolution. 

\begin{figure}
\begin{center}
\includegraphics[scale=0.8]{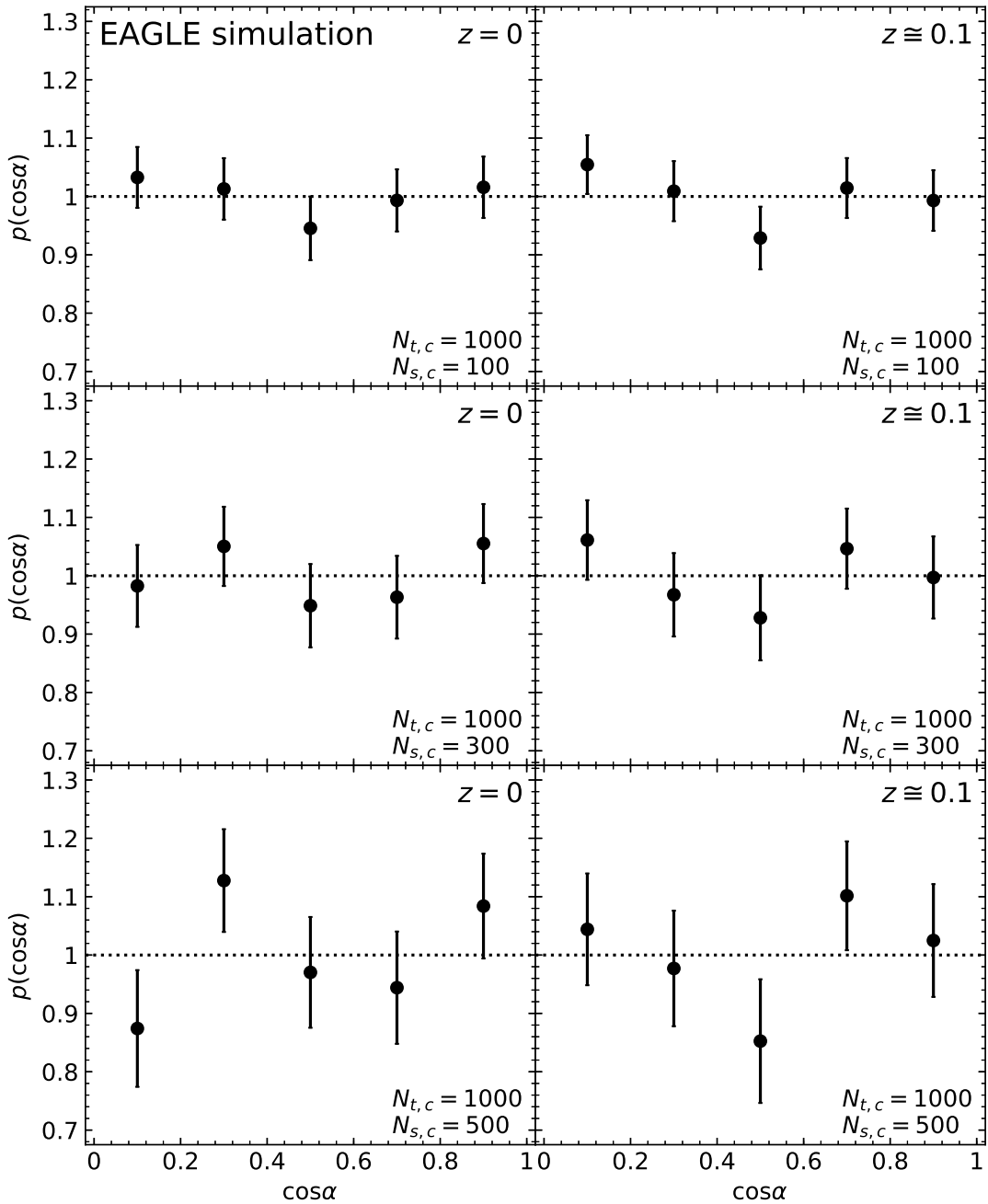}
\caption{Probability density functions of the cosines of the alignment angles between the spin vectors of 
the stellar parts of the subhalos in the isolated pairs from the EAGLE cosmological hydrodynamic simulations 
at two different redshifts for three different cases of the threshold value on the number of the constituent 
particles of a secondary subhalo, $N_{s,c}$. The threshold value on the number of the constituent particles 
of a host halo, $N_{t,c}$, is fixed at $1000$ for all cases. }
\label{fig:pcosa_eagle}
\end{center}
\end{figure}

Each subhalo carries three different angular momentum vectors: One from the DM particles, another from the gas particles, and the 
third from their stellar components. Since the spin axes of the observed spiral galaxies we have determined in Section \ref{sec:ob} 
are the angular momentum vectors of the stellar components but not of the DM nor of the gas particles, we use the third ones to 
measure the intrinsic spin alignments in the isolated pair systems. For each isolated pair system at each redshift, we compute the 
cosine of the angle between the angular momentum vectors of the stellar components of the two subhalos. Unlike the case of the 
observed spiral galaxies, there is no ambiguity in the directions of the spin angular momentum vectors of the stellar components of the 
subhalos. Therefore each isolated halo system yields only one value of the cosine of the alignment angle.  Through the same 
procedures described in Section \ref{sec:ob}, we determine the probability density functions of the cosines of the alignment angles 
between the angular momentum vectors of the stellar components of the primary and the secondary subhalos in the isolated pair 
systems at $z=0$ and $0.1$, which are plotted along with the Poisson errors in the top two panels of Figure \ref{fig:pcosa_eagle}. 

As can be seen, the numerical results at both redshifts are consistent with the uniform distribution (dotted line), showing no signal of 
intrinsic alignment.  To examine the robustness of this result against the numerical resolution, we vary the threshold value, $N_{s,c}$, 
from $100$ to $300$ and to $500$, we remeasure the alignments angles and then redetermine $p(\cos\alpha)$. The middle and bottom 
panels of Figure \ref{fig:pcosa_eagle} show the same as the top panels but for the cases of $N_{s,c}=300$ and $500$, respectively, 
which confirms that the consistency of $p(\cos\alpha)$ with the uniform distribution is robust against the variation of $N_{s,c}$.  

\begin{figure}
\begin{center}
\includegraphics[scale=0.8]{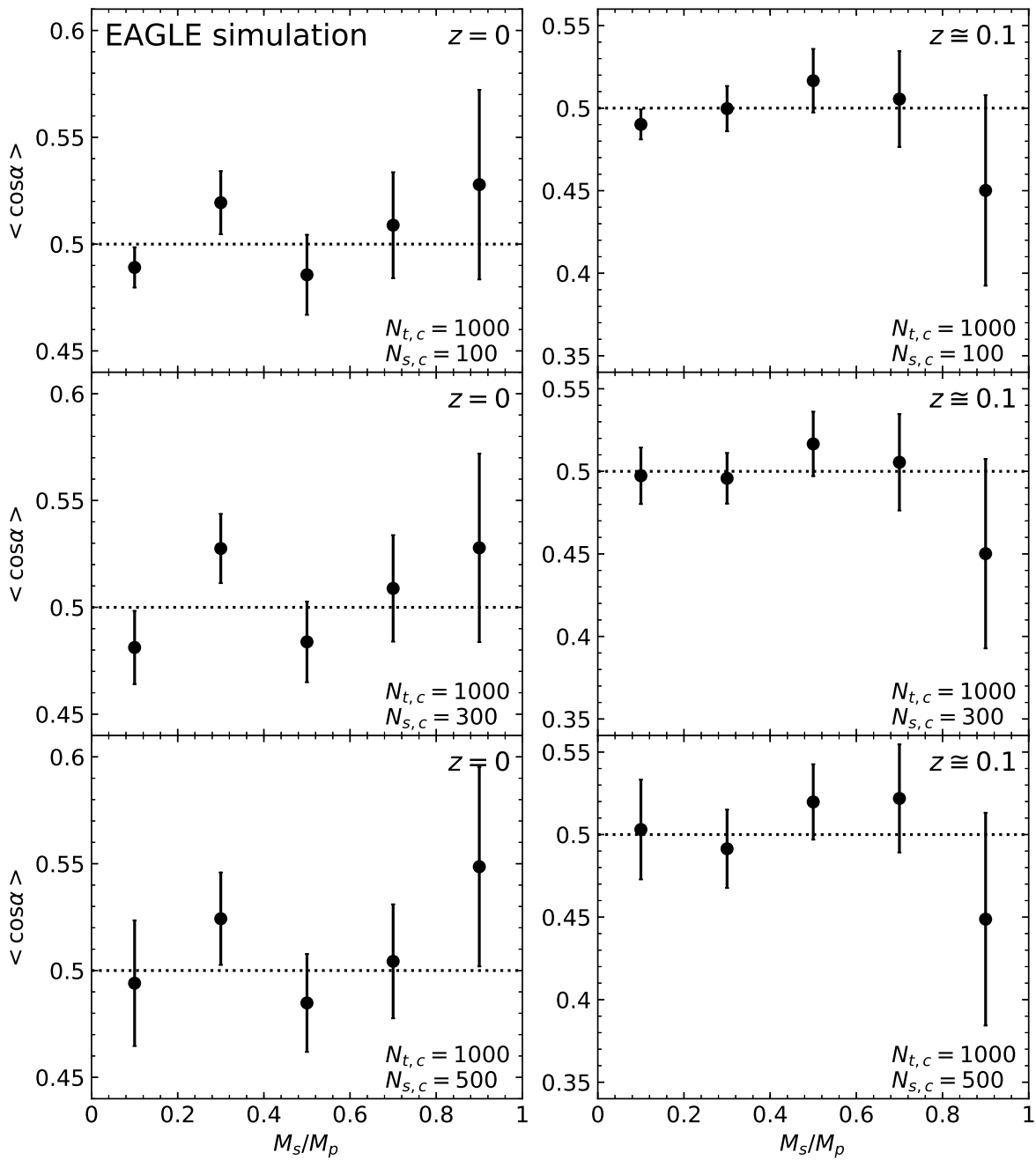}
\caption{Mean values of the cosines of the alignments angles versus the ratios of the masses of the 
secondary subhalos to those of the primary ones in the isolated pair systems from the EAGLE simulations
at two different redshifts for three different cases of $N_{s,c}$.}
\label{fig:mean_mr}
\end{center}
\end{figure}

To answer the question if a signal of intrinsic spin alignment can be found from those pair systems with 
two subhalos having comparable masses, we calculate $\langle\cos\alpha\rangle$ as a function of the ratio of the mass of 
a primary subhalo to that of a secondary subhalo, $M_{s}/M_{p}$, and show the results with the Bootstrap errors 
in Figure \ref{fig:mean_mr}. 
For all of the six cases, $\langle\cos\alpha\rangle$ is found to be uncorrelated with $M_{s}/M_{p}$, revealing that even for those 
pair systems in which two subhalos have comparable masses, the directions of their spin axes are randomly oriented relative 
to each other. 

\begin{figure}
\begin{center}
\includegraphics[scale=0.8]{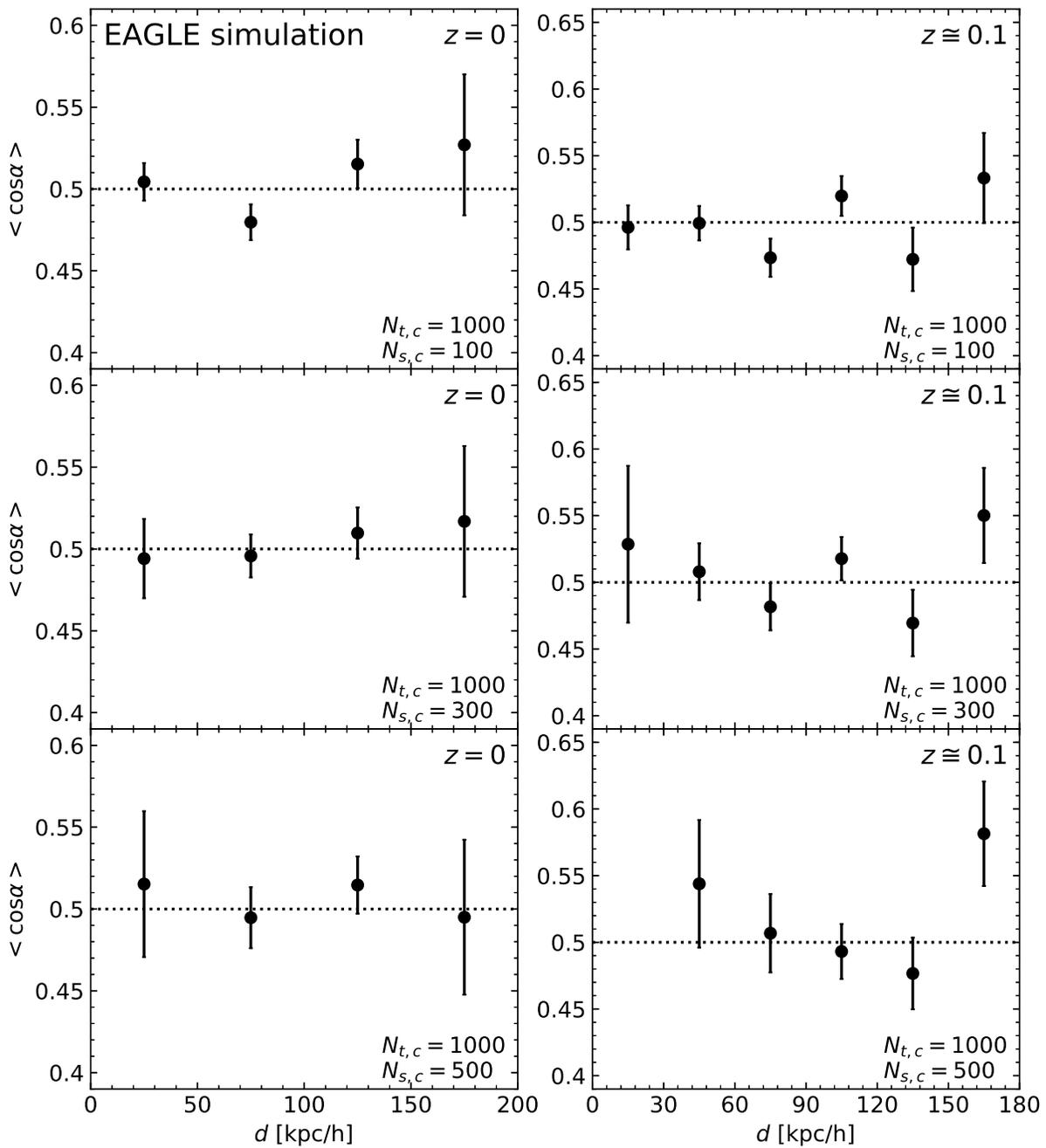}
\caption{Mean values of the cosines of the alignments angles versus the 3D distances between the member 
subhalos in the isolated pair systems from the EAGLE simulations
at two different redshifts for three different cases of $N_{s,c}$.}
\label{fig:mean_3d}
\end{center}
\end{figure}

To see if a signal of intrinsic spin alignment can be found from those pair systems where two subhalos are more closely 
located to each other than the average, we calculate $\langle\cos\alpha\rangle$ as a function of the 3D separation (comoving) 
distances between the subhalos and show the results with the Bootstrap errors in Figure \ref{fig:mean_3d}. As can be seen, the 
average of the cosines of the alignment angles does not show any significant sign of increment with the decrement of the separation 
distance, indicating that no matter how closely located the two subhalos are, their spin axes are not aligned.

\begin{figure}
\begin{center}
\includegraphics[scale=0.8]{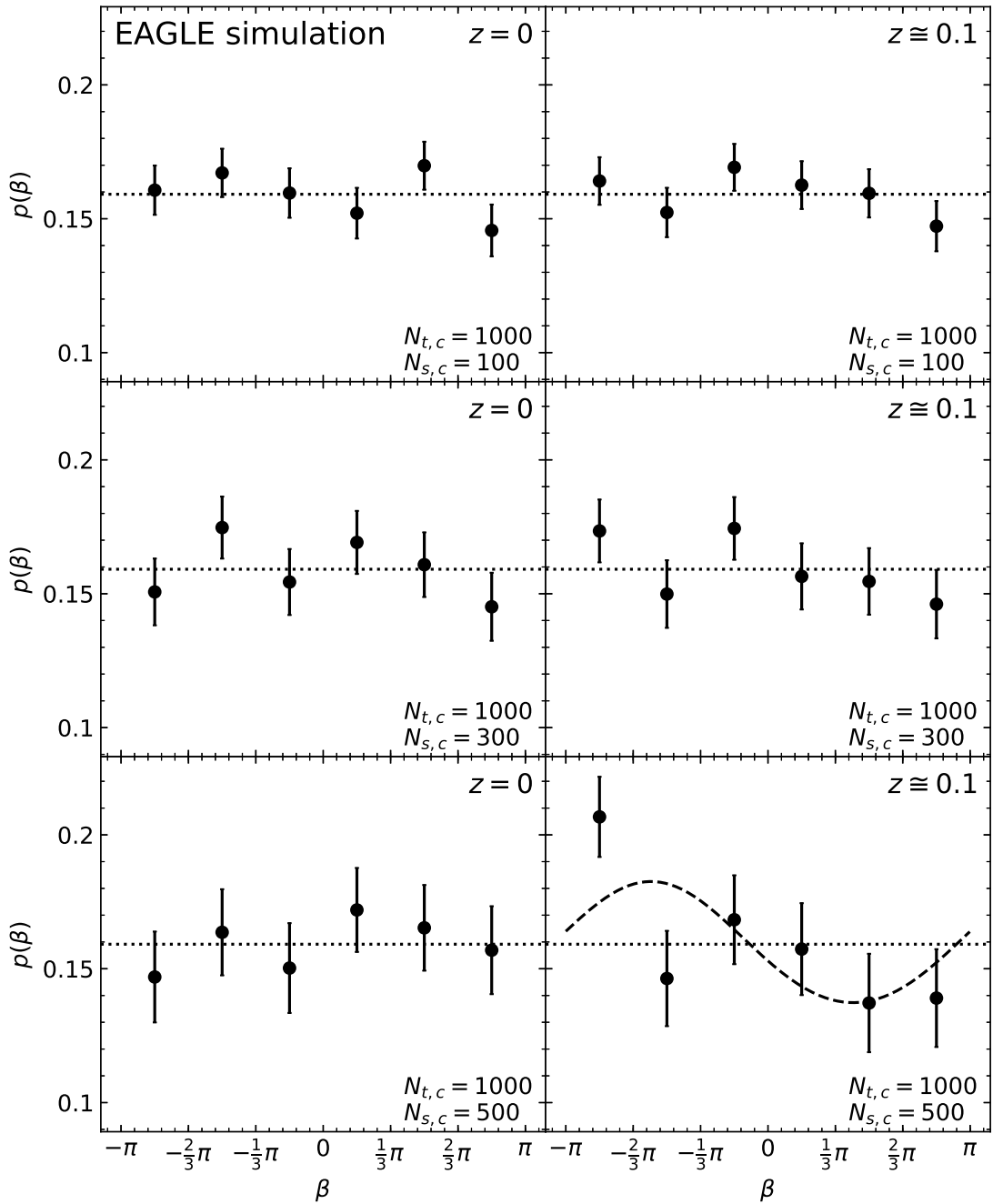}
\caption{Probability density functions of the angles between the 2D projected spin vectors of the stellar parts of 
the subhalos in the isolated subhalo pairs from the EAGLE simulations at two different redshifts each of which has 
three different particle number cuts for three different cases of $N_{s,c}$. 
The uniform distribution is shown for comparison (dotted line) in each panel. The best-fit von Mises distribution 
is shown in the bottom right panel.}
\label{fig:pbeta_eagle}
\end{center}
\end{figure}

As done with the observational data in Section \ref{sec:ob}, it is also investigated whether or not we can find any signal of intrinsic 
alignments between the 2D spin vectors in the isolated pair systems which are obtained by projecting the 3D spin vectors of the 
primary and the secondary subhalos onto the plane perpendicular to the directions of the position vectors of the primary subhalos. 
Measuring the alignment angles between the 2D spin vectors of the primary and the secondary subhalos, we determine the probability 
density functions, $p(\beta)$, and plot them with the Poisson errors in Figure \ref{fig:pbeta_eagle}.   
Although a rather high value of $p(\beta)$ is witnessed at the first bin from the left for the case of $z=0.1,\ N_{s,c}=500$ 
(bottom right panel), all of the other values of $p(\beta)$ seem to agree well with the uniform distribution at both of the redshifts, 
regardless of the threshold values of $N_{s,c}$.  We fit the von Mises distribution to the numerical result for the case of 
$z=0.1,\ N_{s,c}=500$, and show the fitting result as dashed line.  As can be seen, the von Mises distribution does not provide 
a good fit to the numerical result for this case, which implies that the marginally significant deviation of $p(\beta)$ from the uniform 
distribution at the first bin for this case is likely to be a numerical fluke. 
The comparison of Figure \ref{fig:pbeta_eagle} with Figure \ref{fig:pbeta} ensures that the numerical results from the EAGLE 
hydrodynamic simulations is quite consistent with the null hypothesis of no spin alignments in the isolated pair systems. 

\section{Summary and Discussion}\label{sec:dis}

We have investigated whether or not the 3D spin vectors of the member galaxies belonging to the isolated spiral pairs are 
aligned with each other by utilizing the FoF group catalog of the SDSS DR10 galaxies \citep{sdssdr10,tempel-etal14}. 
Defining the isolated spiral pairs as the FoF groups composed of only two spiral (Sab or Scd) galaxies without being embedded in any 
other larger FoF groups, we have detected a clear signal of intrinsic spin alignment between the directions of the 3D spin axes of the 
two member galaxies, showing that the Sab-Sab spiral pairs exhibit the strongest signal. The generalized chi-squared test have 
rejected the null hypothesis of no signal at the $98.37\%$ and $99.99\%$ confidence levels for the case of all and the Sab-Sab spiral 
pairs, respectively.  The strength of the alignment signal has been found to have a mild dependence on the luminosity ratios of the 
member galaxies and their separation distances. A stronger alignment tendency is yielded by the isolated spiral pairs whose member 
galaxies are more closely located to each other and by the pairs whose members have comparable luminosities. 

To complement the parametric chi-squared test operating on the cosines of the angles between the 3D spin vectors, we have employed 
the non-parametric Rayleigh test operating directly on the angles between the 2D spin vectors projected on the plane of sky and 
showed that the null hypothesis is rejected at the $99.9999\%$ confidence level even for the case of all spiral pairs. The comparison of 
the probability density function with the von Mises distribution has determined the best-fit parameters, which again confirmed the 
presence of non-uniformity in the distribution of the angles between the 2D projected spin vectors of the members in the isolated spiral 
pairs. 

To see whether or not the detected signal of intrinsic spin alignment in the isolated spiral pairs can be naturally explained 
by the $\Lambda$CDM cosmology, we have also conducted a similar investigation with the data from the EAGLE cosmological 
hydrodynamic simulations \citep{eagle,eagle_catalog}. Selecting the isolated dark matter halos consisting of only two subhalos from 
the EAGLE cosmological hydrodynamic simulations, we have determined the probability density functions of the alignment angles 
between the spin angular momentum vectors of the stellar parts of two subhalos in each pair. In contrast to the observational result, 
we have failed to find any alignment signal, showing that both of the probability density functions of the cosines of the 3D angles 
and of the 2D angles are quite uniform. We have also found that this numerical result of no alignment tendency is robust against 
varying the threshold values of the particle numbers of the DM halos and the subhalos.
This result agrees well with the previous one obtained by \citet{lee12} who found no signal of alignments between the angular 
momentum vectors of the subhalos in the isolated FoF groups from the H-CoDECS \citep{codecs}.  Recalling that in \citet{lee12} the 
angular momentum vectors of the subhalos were calculated not from the stellar parts but from the DM particles that constitute the 
subhalos, the agreement between the two results implies that for a $\Lambda$CDM cosmology there is no gravitational nor 
hydrodynamical mechanism that can create the alignments between the spin vectors of the subhalos in the isolated pairs.

As the tension between the observational and the numerical results calls for a physical explanation, we bring up three possible 
scenarios. The first scenario concerns about the inaccuracy in the measurements of the directions of the spin axes of the spiral 
galaxies from the SDSS DR10. Since the shapes of real spiral galaxies are not perfectly circular nor infinitesimally thin, the directions 
of their spin axes determined by the circular thin disk approximation may be inaccurate enough to produce systematics in the 
measurements of intrinsic spin alignments in the isolated spiral pairs. 
Furthermore, there is a notable difference between the observational and the numerical analyses. For the former, we 
have considered only the spiral galaxies, excluding the lenticular and ellipticals, simply because the circular thin disk approximation is 
not applicable to the excluded ones.  For the numerical analysis, however, we have considered all isolated pair systems, regardless 
of their morphological types. This difference might be responsible partly for the inconsistency between the observational and the 
numerical results on the intrinsic spin alignments in the isolated pairs. To overcome these difficulties in the measurements of the spin 
directions of the observed galaxies, it will be necessary to develop a more realistic model by which the directions of the spin vectors of 
the spiral, elliptical and lenticular galaxies could be all determined with higher accuracy, which is beyond the scope of this work. 
 
The second scenario ascribes the inconsistency to some missing baryon process that was not included in the EAGLE 
hydrodynamic simulations.  Although the spin vectors of the DM components of the subhalos in the isolated pair systems are not 
aligned with each other as shown by \citet{lee12}, some unknown baryon processes might be able to generate alignments between 
the spin vectors of the stellar counterparts.  If this process were incorporated into a hydrodynamical simulation, then the simulation 
would yield an alignment signal as strong as the detected one from observations.  It will be, however, a daunting task to figure 
out what that missing baryon process should be. 

The third scenario is the most radical one, claiming that the observed signal of intrinsic spin alignments should be a 
new local anomaly that challenges the $\Lambda$CDM model on the small-scale.  The spin alignments in the isolated spiral pairs 
can be produced if the member galaxies interact strongly with each other after they form a bound pair and reside in a dynamically 
isolated state \citep{GT78,hel84,lee12}.  This scenario basically interprets the numerical results from the EAGLE and H-CoDECS 
simulations as an indication that in a $\Lambda$CDM cosmology the mutual interaction between the subhalos do not occur 
efficiently enough to produce the intrinsic alignments, and claims that in the cDE models the fifth force produced by the dark sector 
coupling could stimulate the interactions between the member galaxies in the isolated pairs, generating the intrinsic spin alignments 
\citep{lee12}. 
In this scenario, the stronger signal of intrinsic spin alignment detected from the Sab-Sab pairs may be also explained with the 
same logic: Since the Sab-Sab galaxies are expected to have formed earlier than the Scd-Scd and Scd-Sab pairs,  their members 
must have interacted for a longer period of time during which stronger spin alignments are induced. 
Radical as this third scenario may sound, it would be less difficult to test than the first two, requiring only a 
hydrodynamical simulation for a cDE cosmology.  We plan to explore the above three different scenarios and report the results 
elsewhere in the future.

\acknowledgments

We thank an anonymous referee for providing very helpful suggestions. 
We acknowledge the Virgo Consortium for making their simulation data available. The EAGLE simulations were performed 
using the DiRAC-2 facility at Durham, managed by the ICC, and the PRACE facility Curie based in France at TGCC, CEA, 
Bruy\`{e}res-le-Ch\^{a}tel. This research was supported by Basic Science Research Program through the National Research 
Foundation of Korea(NRF) funded by the Ministry of Education(2016R1D1A1A09918491). It was also partially supported by a 
research grant from the NRF to the Center for Galaxy Evolution Research  (No.2017R1A5A1070354).

\clearpage
\clearpage
\clearpage
\clearpage
\clearpage
\clearpage
\clearpage
\clearpage
\clearpage
\clearpage
\clearpage
\clearpage
\begin{deluxetable}{ccccc}
\tablewidth{0pt}
\setlength{\tabcolsep}{5mm}
\tablecaption{Isolated spiral pairs from the flux-limited sample of the SDSS DR10}
\tablehead{ $z$ & $M_{r}$ & $N_{Sab}$ & $N_{Scd}$ & $N_{Sm}$}
\startdata
$[0.0104,\ 0.1983]$ & $[-22.77,\ -16.02]$ & $1129$ & $523$ & $1369$ 
\enddata
\label{tab:ob}
\end{deluxetable}
\clearpage
\begin{deluxetable}{ccc}
\tablewidth{0pt}
\setlength{\tabcolsep}{5mm}
\tablecaption{Best fit parameters for the von Mises distribution}
\tablehead{Morphology & $\mu$ & $\kappa$ }
\startdata
All & $0.009\pm0.015$ & $0.901\pm0.015$ \\
Sab-Sab & $0.015\pm0.024$ & $0.919\pm0.024$ 
\enddata
\label{tab:best_fit}
\end{deluxetable}
\clearpage
\begin{deluxetable}{cccccc}
\tablewidth{0pt}
\setlength{\tabcolsep}{5mm}
\tablecaption{Isolated pair systems from the EAGLE project}
\tablehead{$z$ & $N_{t,c}$ & $N_{s,c}$ & $M_{p}$ & $M_{s}$ & $N_{pair}$ \\
& & & ($10^{8}\,h^{-1}M_{\odot}$) & ($10^{8}\,h^{-1}M_{\odot}$) & }
\startdata
$0$ & 1000 & 100 & $[53.4,\ 2.97\times 10^{3}]$ & $[7.93,\ 3.98\times 10^{2}]$ & $1777$ \\
$0$   & 1000 & 300 & $[53.4,\ 2.97\times 10^{3}]$ & $[26.6,\ 3.98\times 10^{2}]$ & $1033$ \\
$0$    & 1000 & 500 & $[55.1,\ 1.60\times 10^{3}]$ & $[42.8,\ 3.98\times 10^{2}]$ & $572$ \\
$0.1$ & 1000 & 100 & $[51.1,\ 2.74\times 10^{3}]$ & $[6.06,\ 3.01\times 10^{2}]$ & $1868$ \\
$0.1$     & 1000 & 300 & $[51.1,\ 2.74\times 10^{3}]$ & $[24.9,\ 3.01\times 10^{2}]$ & $1013$ \\
$0.1$      & 1000 & 500 & $[51.1,\ 1.72\times 10^{3}]$ & $[44.2,\ 3.01\times 10^{2}]$ & $522$
\enddata
\label{tab:eagle}
\end{deluxetable}

\end{document}